\documentstyle[
               twocolumn,
               epsfig,amssymb,aps]{revtex}
 
\newcommand{\etal}{{\em et al.}}
\newcommand{\rms}{r_{\rm nn}^{\rm rms}}
 
\begin{document}
 
 
\title{Three-body correlations in Borromean halo nuclei}
 
\author{F.M.~Marqu\'{e}s$^a$\thanks{e-mail: {\tt Marques@caelav.in2p3.fr}},
M.~Labiche$^a$\thanks{Present address: University of Paisley, Scotland.},
N.A.~Orr$^a$, J.C.~Ang\'elique$^a$, L.~Axelsson$^b$, B.~Benoit$^c$,
U.C.~Bergmann$^d$, M.J.G.~Borge$^e$, W.N.~Catford$^f$, S.P.G.~Chappell$^g$,
N.M.~Clarke$^h$, G.~Costa$^i$,
N.~Curtis$^f$\thanks{Present address: University of Birmingham, UK.},
A.~D'Arrigo$^c$, E.~de~G\'{o}es~Brennand$^c$, F.~de~Oliveira~Santos$^j$,
O.~Dorvaux$^i$, G.~Fazio$^k$, M.~Freer$^{h,a}$, B.R.~Fulton$^h$,
G.~Giardina$^k$, S.~Gr\'evy$^l$\thanks{Present address: LPC, Caen, France.},
D.~Guillemaud-Mueller$^l$, F.~Hanappe$^c$, B.~Heusch$^i$, B.~Jonson$^b$,
C.~Le~Brun$^a$\thanks{Present address: ISN, Grenoble, France.},
S.~Leenhardt$^l$, M.~Lewitowicz$^j$,
M.J.~L\'opez$^j$\thanks{Present address: CENBG, Bordeaux, France.},
K.~Markenroth$^b$, A.C.~Mueller$^l$,
T.~Nilsson$^b$\thanks{Present address: ISOLDE, CERN, Switzerland.},
A.~Ninane$^a$\thanks{On leave from: UCL, Louvain-la-Neuve, Belgium.},
G.~Nyman$^b$, I.~Piqueras$^e$, K.~Riisager$^d$, M.G.~Saint~Laurent$^j$,
F.~Sarazin$^j$\thanks{Present address: University of Edinburgh, Scotland.},
S.M.~Singer$^h$, O.~Sorlin$^l$, L.~Stuttg\'e$^i$}
 
\address{$^a$ Laboratoire de Physique Corpusculaire,
IN2P3-CNRS, ISMRA et Universit\'e de Caen, F-14050 Caen cedex, France}
\address{$^b$ Experimentell Fysik, Chalmers Tekniska H\"ogskola, S-412 96
G\"{o}teborg, Sweden}
\address{$^c$ Universit\'e Libre de Bruxelles, CP 226,
B-1050 Bruxelles, Belgium}
\address{$^d$ Institut for Fysik og Astronomi, 
 Aarhus Universitet, DK-8000 Aarhus C, Denmark}
\address{$^e$ Instituto de Estructura de la Materia, CSIC,
E-28006 Madrid, Spain}
\address{$^f$ Department of Physics, University of Surrey,
Guildford, Surrey, GU2~7XH, U.K.}
\address{$^g$ Department of Nuclear Physics, University of
Oxford, Keble Road, Oxford OX1 3RH, U.K.}
\address{$^h$ School of Physics and Astronomy, University of Birmingham,
Birmingham B15 2TT, U.K.}
\address{$^i$ Institut de Recherche Subatomique, IN2P3-CNRS,
Universit\'e Louis Pasteur, BP 28, F-67037 Strasbourg cedex, France}
\address{$^j$ GANIL, CEA/DSM-CNRS/IN2P3, BP 55027, F-14076 Caen cedex, France}
\address{$^k$ Dipartimento di Fisica, Universit\`a di Messina,
Salita Sperone 31, I-98166 Messina, Italy}
\address{$^l$ Institut de Physique Nucl\'eaire, IN2P3-CNRS,
F-91406 Orsay cedex, France}
 
\date{\today}
 
\maketitle
 
\begin{abstract}
Three-body correlations in the dissociation of two-neutron halo nuclei
are explored using a technique based on intensity interferometry and Dalitz
plots. This provides for the combined treatment of both the n-n and core-n 
interactions in the exit channel. As an example, the breakup of $^{14}$Be into 
$^{12}$Be+n+n by Pb and C targets has been analysed and the halo n-n separation 
extracted. A finite delay between the emission of the neutrons in the reaction 
on the C target was observed and is attributed to $^{13}$Be resonances 
populated in sequential breakup.
\end{abstract}
 
 
The quest for the drip-lines, which define the limits of binding for nuclear
systems, has only been attained for light nuclei. As such, these nuclei are
unique in displaying the manner in which nucleons bind in an $A$ nucleon system
from the most neutron deficient to the most neutron rich. Clustering phenomena,
observed for example recently in excited states close to threshold
\cite{Fre99}, also appear as haloes in ground states near the neutron drip-line
\cite{Han87}. The most intriguing manifestation of clustering are the Borromean
two-neutron halo nuclei ($^{6}$He, $^{11}$Li and $^{14}$Be), in which the
two-body subsystems are unbound \cite{Zhu93}. Beyond the global properties,
such as the abnormal sizes \cite{Tan85,AlK96} or the low momentum content of
the constituents \cite{NAO97}, the experimental challenge lies in determining
the structure of these three-body systems.
 
The dissociation in the field of a target nucleus, followed by the measurement
of the momenta of the fragments (core+n+n), has been used in attempts to probe
correlations in two-neutron halo nuclei. Vestiges, however, of the two-body
forces that stabilise the projectile in the ground state may affect the
three-particle decay in the form of final-state interactions (FSI).
Experimentally, beyond the reconstruction of the core+2n invariant mass, the 
analyses so far reported have been restricted to the binary channels. The 
relative energy in the core+n channel has been used, for example, to probe the 
formation of core-n resonances \cite{Zin97,Ale98,Aum99}. The n-n observables, 
which should provide access to correlations within the halo, have often been 
compared only to simplified interpretations, such as a di-neutron configuration 
or three-body phase space \cite{Zin97,Sac93}, neglecting the n-n FSI.
 
In the present paper the three-body correlations in the dissociation of
two-neutron halo nuclei are explored. In particular, a new method for analysing
triple coincidence events (core+n+n) from kinematically complete experiments is
described. The method incorporates the techniques of intensity interferometry
\cite{Boa90} and Dalitz plots \cite{Per87} and permits the halo n-n separation
and time delay between the emission of the two neutrons to be derived. As will
be seen, the latter is related to the presence of core-n FSI in the exit
channel. In principle, the present approach also allows the energies and 
lifetimes of these resonances (or virtual states if $\ell=0$) to be derived.
 
When neutrons are emitted in close proximity in space-time, the wave function
of relative motion is modified by the known FSI and quantum statistical
symmetries \cite{Boa90}. Two-neutron intensity interferometry, and in
particular the correlation function \cite{Kop74} $C_{\rm{nn}}$, relates this
modification to the space-time separation of the particles at emission:
\begin{equation}
 C_{\rm{nn}}(p_1,p_2)=\frac{d^2n/dp_1dp_2}{(dn/dp_1)\,(dn/dp_2)} \label{e:C12}
\end{equation}
The numerator is the measured two-particle distribution and the denominator the 
product of the independent single-particle distributions \cite{Kop74,FMM00}, 
which are commonly projected onto one dimension, the relative momentum 
$q=|\vec{p}_1-\vec{p}_2|$. In an earlier paper \cite{FMM00}, the correlation
function was extracted for the dissociation of $^{14}$Be by Pb with the aim of
probing the spatial configuration of the halo neutrons. The analysis followed
the formalism of Ref.~\cite{Led82} and assumed that the neutrons were emitted
simultaneously following dissociation in the Coulomb field of the target. A n-n
separation of $\rms=5.6\pm1.0$~fm was thus obtained.
 
\begin{figure}[tb]
 \begin{center}
  \mbox{\psfig{file=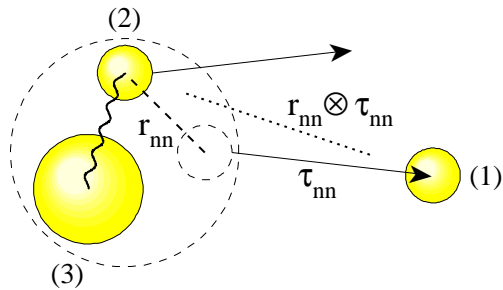,width=8.5cm}}
 \end{center}
 \caption{Schematic view of the sequential breakup of a two-neutron halo
nucleus, whereby dissociation passes through a resonance in the core-n system.
A delay $\tau_{\rm{nn}}$, equivalent to the lifetime of the resonance, is
introduced in the emission of the second neutron. The effective source size is
a convolution of the n-n separation $r_{\rm{nn}}$ and the delay 
$\tau_{\rm{nn}}$.} \label{f:rt0}
\end{figure}

The same analysis has been applied to dissociation of $^{14}$Be by a C target,
in order to investigate the influence of the reaction mechanism \cite{FMM--}. 
The n-n separation obtained was somewhat larger, $\rms=7.6\pm1.7$~fm. This 
raises the question as to whether simultaneous emission can be assumed a
priori. In principle, the analysis of the correlation function in two
dimensions, transverse and parallel to the total momentum of the pair, would
allow for the unfolding of the source size and lifetime \cite{Led82}. Such an
analysis requires a large data set and was thus not applicable to the present
measurements. The system being studied here is far less complex, however, than
those usually encountered in interferometry (for example compound nuclei
evaporating particles or systems of colliding heavy ions \cite{Boa90}).
Moreover, the simple three-body nature of the system breaking up suggests
immediately that any delay in the emission of one of the neutrons will arise
from core-n FSI/resonances in the exit channel (Fig.~\ref{f:rt0}). As will be
seen later, the degree to which such resonances are present may depend on the
reaction mechanism. 
 
The data examined here were acquired from the dissociation of a 35~MeV/N
$^{14}$Be beam into $^{12}$Be+n+n by Pb and C targets. Details of the
experiment and previous analyses have been reported elsewhere
\cite{FMM00,Lab--}. Each event was reconstructed from the momenta
$\vec{p}_{1,2}$ of the neutrons measured using the DEMON array \cite{fmm00} as
follows: (i) we calculate the average velocity $\langle\beta\rangle$ of the
core+n+n frame at dissociation as that for which the mean for all events of the
total neutron momentum along the beam axis
$\langle\vec{p}_1+\vec{p}_2\rangle_z$ is 0 and the average decay energy of the
system $\langle{E_{\rm{d}}}\rangle$ (see below) is a minimum; (ii) in this
frame, momentum conservation is applied event-by-event to reconstruct the core
momentum $\vec{p}_3=-(\vec{p}_1+\vec{p}_2)$. From the four-momenta of the three
particles we calculate the total energy available in the center-of-mass of the
system and extract the kinetic part, the decay energy:
\begin{equation}
 E_{\rm{d}} = \sqrt{\left(\sum_{i=1}^3p_i\right)^2}-\sum_{i=1}^3m_i
\end{equation}
Note that in the present analysis the decay has been reconstructed from only
the neutron momenta $\vec{p}_{1,2}$. The use of this reconstruction algorithm
instead of the more classical analysis employing the core momentum $\vec{p}_3$
measured in the charged-particle telescope eliminates the limited energy and
position resolution of this detector and the uncertainty related to the depth
within the target at which the reaction took place. Importantly, the form of
the spectra obtained here for the $^{14}$Be and core-n invariant masses agree
well with those obtained using the core momentum \cite{Lab--,Jon--}.
 
An interacting phase-space model has been developed for the analysis of triple
correlations in the data. In brief, the experimental decay energy distribution
is used as input to generate events $\vec{p}_{1,2,3}(E_{\rm{d}})$ following
three-body phase space \cite{Nik68}. The core-n resonances are introduced 
following the sequential breakup of the system (Fig.~\ref{f:rt0}) into one 
neutron and the core-n resonance with a relative energy $E_{23}$ given by a 
Breit-Wigner distribution $(E_0,\Gamma)$; the resonance is then allowed to 
decay into the core plus neutron. In the n-n channel, the FSI is introduced via 
a probability $P(|\vec{p}_1-\vec{p}_2|)$ to accept the event following the form 
of the measured n-n correlation function \cite{FMM00,Led82,FMM--}. The final 
momenta $\vec{p}_{1,2,3}$ are filtered through a simulation including all 
experimental effects \cite{FMM00,fmm00} and the reconstruction algorithm 
described above is applied to $\vec{p}_{1,2}$.
 
\begin{figure}[tb]
 \begin{center}
  \mbox{\psfig{file=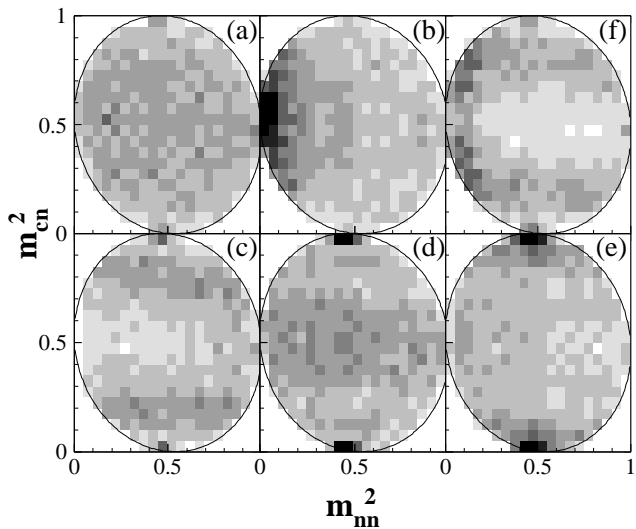,width=8.5cm}}
 \end{center}
 \caption{Dalitz plots (core-n versus n-n) for simulations with the interacting
phase-space model of $^{14}$Be dissociation: without FSI (a), with n-n FSI (b)
and with a core-n resonance at $E_0=0.8$ (c), 2.0 (d) and 3.5~MeV (e). The
combination of the n-n and the core-n FSI of (c) is shown in (f).} \label{f:sim}
\end{figure}

\begin{figure}[tb]
 \begin{center}
  \mbox{\psfig{file=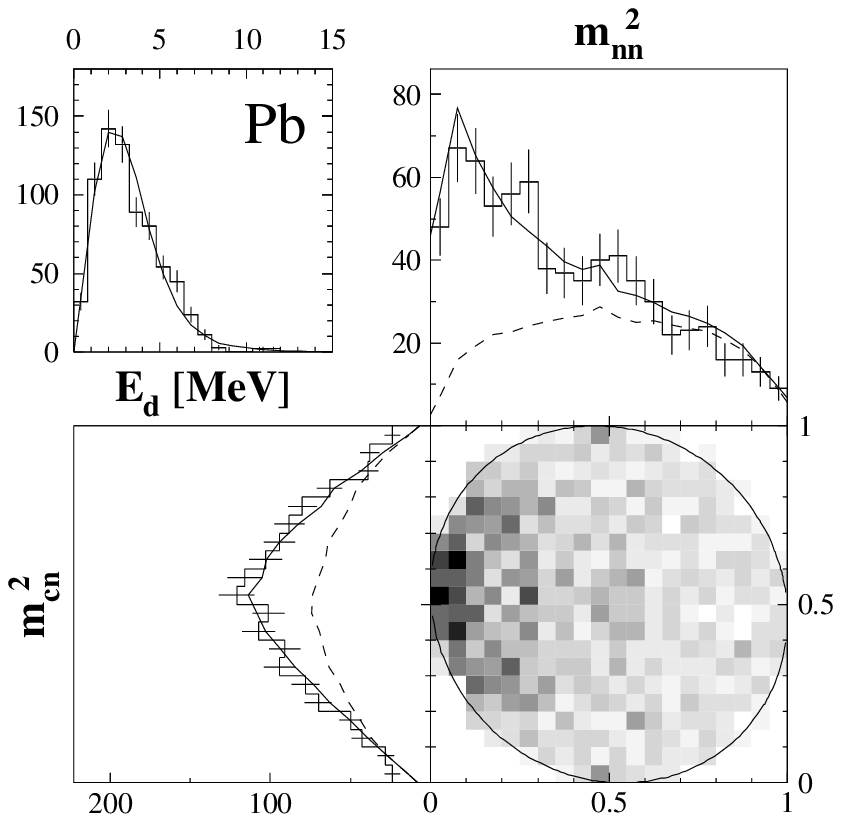,width=8.5cm}}
  \mbox{\psfig{file=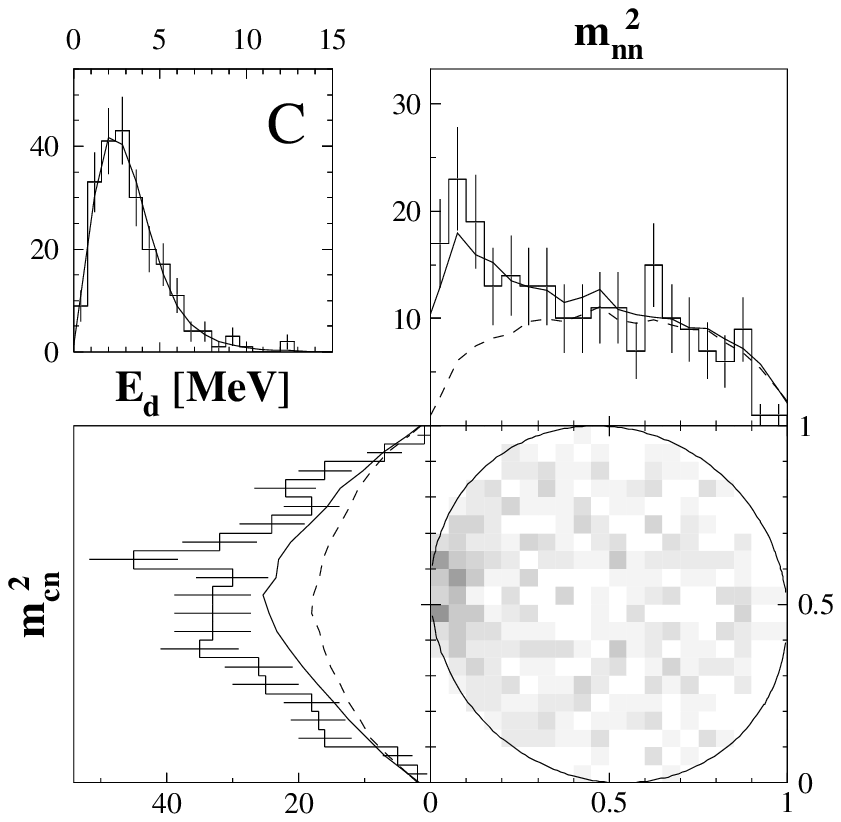,width=8.5cm}}
 \end{center}
 \caption{Dalitz plots (core-n versus n-n), and the projections onto both axes,
for the data from the dissociation of $^{14}$Be by Pb (upper) and C (lower
panels). The lines are the results of the phase-space model simulations
with/without (solid/dashed) n-n FSI. The insets correspond to the $^{14}$Be
decay energy.} \label{f:ps2}
\end{figure}

\begin{figure}[tb]
 \begin{center}
  \mbox{\psfig{file=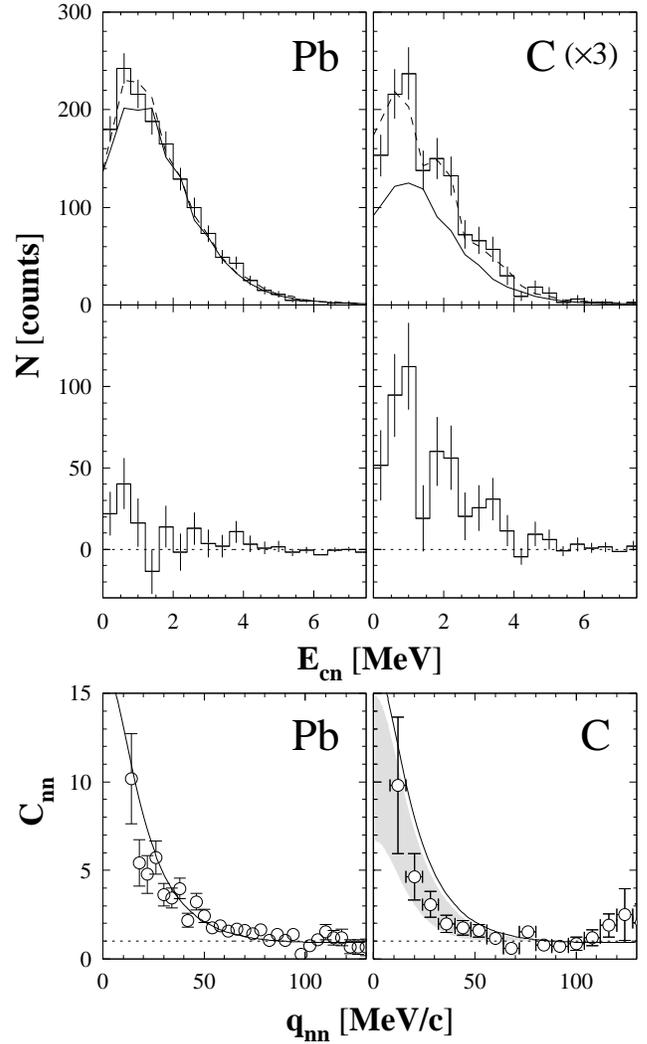,width=8.5cm}}
 \end{center}
 \caption{Core-n relative energy distributions (upper) and n-n correlation 
functions (lower panels) for the dissociation of $^{14}$Be by Pb and C. The 
lines in the $E_{\rm{cn}}$ spectra are the result of the phase-space model
simulations with n-n FSI (solid) plus core-n resonances (dashed, see text). The
histograms presented in the middle panels are the difference between the data 
and the n-n FSI simulations. The solid lines in the lower panels are the 
$C_{\rm{nn}}$ for $\rms=5.6$~fm and $\tau_{\rm{nn}}=0$; the shaded area 
corresponds to the inclusion of a finite $\tau_{\rm{nn}}$ (see text).}
 \label{f:bec}
\end{figure}

Correlations in three-particle decays have been extensively studied in particle
physics by means of Dalitz plots of the particle energies $(E_i,E_j)$ or the
squared invariant masses of particle pairs $(M_{ij}^2,M_{ik}^2)$, with
$M_{ij}^2=(p_i+p_j)^2$. In these representations, FSI/resonances lead to a
non-uniform population of the surface within the kinematic boundary defined by
energy-momentum conservation and the decay energy \cite{Per87}. The classic
example of such an analysis is the three-body decay of an unstable particle
\cite{Ait01}.
In the present case, the core+n+n system exhibits a distribution of decay
energies. Consequently, the value of $E_{\rm{d}}$ associated with each event
will lead to a different boundary for the Dalitz plot, and the resulting plot
containing all events cannot be easily interpreted. We have thus introduced a
normalized invariant mass:
\begin{equation}
 m_{ij}^2 = \frac{M_{ij}^2-(m_i+m_j)^2}{(m_i+m_j+E_{\rm{d}})^2-(m_i+m_j)^2}
\end{equation}
which ranges between 0 and 1 ($E_{ij}$ between 0 and $E_{\rm{d}}$) for all 
events and exhibits a single kinematic boundary. Examples of how n-n and core-n 
FSI present in the decay are manifested in core-n versus n-n Dalitz plots are
displayed in Fig.~\ref{f:sim}, whereby events have been simulated with the
model described above. The inputs were an $E_{\rm{d}}$ distribution following
that measured (Fig.~\ref{f:ps2}), the $C_{\rm{nn}}$ obtained with the C target
\cite{FMM--} (Fig.~\ref{f:bec}), and core-n resonances with $\Gamma=0.3$~MeV at
$E_0=0.8,2.0,3.5$~MeV. 
 
In the absence of any FSI, the Dalitz plot exhibits, as noted above, a uniform
population (Fig.~\ref{f:sim}a). The n-n FSI appears as a concentration of 
events with $m_{\rm{nn}}^2\lesssim0.25$ (Fig.~\ref{f:sim}b), which correspond 
to small relative momenta \cite{FMM00,FMM--}. The core-n resonance at
$E_0=0.8$~MeV (Fig.~\ref{f:sim}c) appears as horizontal bands around
$m_{\rm{cn}}^2\approx0.25$ and $0.75$. The location of these bands depends on 
the energy of the resonance with respect to the mean decay energy of the 
system: a single band at $m_{\rm{cn}}^2\approx0.5$ if 
$E_0\sim\langle{E_{\rm{d}}}\rangle$ (Fig.~\ref{f:sim}d) and two symmetric bands
if $E_0\gtrless\langle{E_{\rm{d}}}\rangle$ (Fig.~\ref{f:sim}c,e) \cite{FMM--}.
This feature arises as when one of the neutrons forms a resonance with the core 
at a given value of $m_{23}^2$, the relative core-n energy of the other is 
essentially fixed at $m_{13}^2\approx1-m_{23}^2$ \cite{FMM--}. Importantly, the 
Dalitz plot representation provides not only for the identification of the 
different FSI between the particle pairs, but also a direct comparison of the 
relative importance of each in the decay (Fig.~\ref{f:sim}f).

The Dalitz plot for the data from the dissociation by Pb (Fig.~\ref{f:ps2})
presents a strong n-n FSI and a uniform density for $m_{\rm{nn}}^2\gtrsim0.5$.
Indeed, the n-n FSI alone describes very well the projections onto both axes,
and therefore suggests that core-n resonances are not present to any
significant extent. This result confirms the hypothesis of simultaneous n-n
emission employed in the original analysis of the dissociation of $^{14}$Be by 
Pb \cite{FMM00}. The value of $\rms$ so extracted, $5.6\pm1.0$~fm, thus
corresponds to the n-n separation in the halo of $^{14}$Be.
 
For dissociation by the C target (Fig.~\ref{f:ps2}), despite the lower
statistics, two differences are evident. Firstly, the n-n signal is weaker,
indicating, as discussed earlier, that a significant delay has occurred between
the emission of each neutron. Second, and more importantly, the agreement
between the model including only the n-n FSI and the data for $m_{\rm{cn}}^2$
is rather poor. In order to verify whether this disagreement corresponds to the
presence of core-n resonances, which would be responsible for the weakening of
the n-n signal, we have investigated the core-n relative energy, $E_{\rm{cn}}$.
It has been reconstructed for the simulations incorporating only the n-n FSI
and compared in Fig.~\ref{f:bec} to the data (the model calculations have been
normalized to the data above 4~MeV). For dissociation by Pb, the inclusion of
only the n-n FSI provides a very good description of the data, with the
exception of small deviations below 1~MeV. This is in line with the Dalitz plot 
analysis discussed above.
 
The deviations observed for the C target between the measured $m_{\rm{cn}}^2$
and the simulation including only the n-n FSI (Fig.~\ref{f:ps2}) clearly
correspond to structures in the $E_{\rm{cn}}$ spectrum. Moreover, these
structures are located at energies that are in line with those of states
previously reported in $^{13}$Be: the supposed $d_{5/2}$ resonance at 2.0~MeV
\cite{Bel98} and a lower-lying state(s) \cite{Jon--,Bel98,Tho98}, also
suggested by various theoretical calculations \cite{Zhu95,Des95,Tho99,Lab99}.
The model-to-data ratio is about 1/2, indicating that the peaks correspond to
resonances formed by one of the neutrons in almost all decays; the solid line 
in Fig.~\ref{f:bec} accounts for the contribution of the neutron not 
interacting with the core. If we add to the phase-space model with n-n FSI
core-n resonances ($\Gamma=0.3$~MeV) for all events at $E_0=0.8$, 2.0
\cite{Bel98} and 3.5~MeV\footnote{The present data are not particularly
sensitive to the location and form of the states, in particular below 1~MeV, 
and a resonance at 0.5~MeV would, for example, equally well describe the data.} 
with intensities of 45, 35 and 20\%, respectively, the data are well reproduced 
(dashed line). In the case of dissociation by Pb, the lowest-lying resonance(s) 
appears to be present in at most 10\% of events.
 
The different results obtained for the Pb and C targets may be attributed to
the associated reaction mechanisms \cite{Gar00}. In the case of the Pb target,
the dominant process is electromagnetic dissociation \cite{Lab--}, whereby the
halo neutrons behave as spectators and only the charged core is acted on by the
Coulomb field of the target \cite{Han98}. Qualitatively then, the n-n FSI may
be expected to influence most strongly the decay. In the case of the C target,
nuclear breakup dominates and the reaction takes place at smaller impact
parameters, in general through the interaction of one of the halo neutrons with
the target \cite{Lab--}. As such the population of core-n resonances is
favoured \cite{Ale98,Aum99}.
 
By combining the information extracted from the core-n channel with the n-n
correlation functions, we can extend the analysis and also extract the average
lifetime of the core-n resonances. If we fix the n-n distance in $^{14}$Be as
that obtained for dissociation by Pb, $\rms=5.6$~fm, we can introduce the delay
between the emission of the neutrons $\tau_{\rm{nn}}$ needed to describe the
n-n correlation function for the C target. As discussed earlier, this delay
should correspond to the lifetime of the resonances (Fig.~\ref{f:rt0}). The
result, shown as the shaded area in Fig.~\ref{f:bec}, suggests an average
lifetime of $150^{+200}_{-100}$~fm/$c$, or $(5^{+7}_{-3})\times10^{-22}$~s.
 
In summary, three-body correlations in the dissociation of two-neutron halo
nuclei have been explored. A new analysis technique employing intensity
interferometry and Dalitz plots has been presented and applied to the breakup
of $^{14}$Be by Pb and C targets. Through the combined treatment of both the
n-n and core-n correlations, the halo n-n separation has been extracted and a
finite delay found between the emission of the neutrons for the reaction on C.
This delay can be attributed to resonances in $^{13}$Be populated in sequential
breakup. The application of the techniques presented here to a well established
system such as $^6$He would be of particular interest, as would the
investigation of multi-neutron haloes. Finally, as the technique of intensity
interferometry is applicable to protons, proton-rich nuclei exhibiting similar
few-body clustering may also be explored.
 
The support provided by the staffs of LPC and GANIL in preparing and executing
the experiments is gratefully acknowledged. This work was funded under the
auspices of the IN2P3-CNRS (France) and EPSRC (United Kingdom). Additional
support from the ALLIANCE programme (Minist\`ere des Affaires Etrang\`eres and
British Council), the Human Capital and Mobility Programme of the European
Community (contract n$^\circ$ CHGE-CT94-0056) and the GDR Noyaux Exotiques
(CNRS-CEA) is also acknowledged.

\end{document}